\documentclass[a4paper,12pt]{article}

\usepackage{amsmath}
\usepackage[final]{graphicx}
\usepackage{bm}
\usepackage{amssymb}

\begin{document}
\hfill
\begin{minipage}{20ex}\small
ZTF-EP-15-02
\end{minipage}

\begin{center}
\baselineskip=2\baselineskip
\textbf{\LARGE{Scotogenic R\boldmath$\nu$MDM\\ 
at Three-Loop Level}}\\[6ex]
\baselineskip=0.5\baselineskip

{\large Petar~\v{C}uljak,
Kre\v{s}imir~Kumeri\v{c}ki and
Ivica~Picek
}\\[4ex]
\begin{flushleft}
\it
Department of Physics, Faculty of Science, University of Zagreb,
 P.O.B. 331, HR-10002 Zagreb, Croatia\\[3ex]
\end{flushleft}
\today \\[5ex]
\end{center}

\begin{abstract}

We propose a model in which the radiative neutrino (R$\nu$) masses are induced by 
fermion quintuplet and scalar septuplet fields from the  minimal-dark-matter (MDM) setup.
In conjunction with the 2HDM fields, on top of which our model is built, these hypercharge zero  
fields and additional scalar quintuplet lead to an accidental DM-protecting $Z_2$ symmetry and establish
the R$\nu$MDM model at the three-loop level. 
We assess the potential for discovery of quintuplet fermions on
present and future $pp$ colliders.
\end{abstract}

\vspace*{2 ex}

\begin{flushleft}
\end{flushleft}

\clearpage

\section{Introduction}

After the discovery of $\sim$125 GeV particle~\cite{Aad:2012tfa,Chatrchyan:2012ufa} looking very much like the one Higgs boson of the Standard Model (SM),
searching for the dark matter (DM) is one of the main targets for the next run of the LHC. At the same time, the scotogenic models of radiative neutrino mass
relate the experimental evidence of neutrino masses and the existence of the DM in the Universe~\cite{Agashe:2014kda}.
The scotogenic one-loop model proposed by Ma~\cite{Ma:2006km} augments the SM particle content by three singlet Majorana fermions and a second scalar doublet, and
remains the simplest scotogenic realization. It imposes an exactly conserved $Z_2$ symmetry to both eliminate the unwanted  Yukawa interactions responsible for
the tree-level neutrino masses
and to simultaneously stabilize the DM candidates.
Here we take under scrutiny the notions of (i) {\em loop-generated} neutrino masses and (ii) $Z_2$ {\em symmetry}, which have since then become a common theme among many studies.

A three-loop neutrino masses under consideration  bear an unquestionable appeal of naturally explaining the twelve orders of magnitude hierarchy 
between neutrino masses and the electroweak scale.
The other theme we address here is how to avoid a common use of an {\em ad hoc} $Z_2$  symmetry. In a recent proposal~\cite{Ma:2013yga} to promote  $Z_2$ 
to a local gauge $U(1)_D$ symmetry in a model with two dark scalar doublets transforming as $\pm1$ under $U(1)_D$, 
a breaking of gauge symmetry provides the dynamical origin of an exact $Z_2$ symmetry.
Alternatively, the DM protecting $Z_2$ symmetry may arise ``accidentally'',  on account of the SM symmetry 
and a choice of the field content\footnote{In a recent scotogenic variant realized by a real triplet scalar field~\cite{Brdar:2013iea} 
there is  no need to eliminate the tree-level contribution, 
but additional  $Z_2$ symmetry is needed to stabilize a DM candidate. Also, an antisymmetric tensor field may be stable~\cite{Cata:2014sta} without introducing a new $Z_2$ symmetry.}. 
A realization which is in our focus here is enabled by employing higher weak multiplets, studied within the minimal dark matter (MDM) model~\cite{Cirelli:2005uq}. There, 
an isolated fermion quintuplet $\Sigma \sim (5,0)$  or an isolated scalar septuplet $\chi \sim (7,0)$ have been selected to provide a viable DM candidate.
Recent minimalistic variant of MDM model, with wino-like fermion triplet~\cite{Cirelli:2014dsa}, relies on  the enforcement of accidental B-L symmetry for DM stability.

In the present study we reconsider original MDM multiplets in such a way that both fermion quintuplet and scalar septuplet have to be employed 
to produce the neutrino masses at three-loop level.
In the original model of radiative neutrino masses with MDM (R$\nu$MDM~\cite{Cai:2011qr}) aiming at an automatic $Z_2$ symmetry by 
employing the higher multiplets,
the zero hypercharge quintuplet $\Sigma \sim (5,0)$ has been accompanied by a hypercharge-one sextuplet  $\Phi \sim (6,1)$
to close a one-loop neutrino mass diagram. 
However, as observed in~\cite{Kumericki:2012bf},  there is an additional renormalizable quartic term in this model which 
violates the $Z_2$ symmetry and threatens the stability of the proposed DM candidate.

In the present account we attempt to restore the R$\nu$MDM idea in a three-loop variant which employs both 
the fermion quintuplet $\Sigma \sim (5,0)$ and the scalar septuplet $\chi \sim (7,0)$.  
This is in contrast to an early three-loop model proposed by Krauss, Nasri and Trodden (KNT)~\cite{Krauss:2002px}
followed by similar recent attempts to explain small neutrino masses by adding to the SM content only additional weak singlets~\cite{Hatanaka:2014tba}, 
or by substituting a real scalar triplet for a charged scalar singlet in the KNT model~\cite{Jin:2015cla}.
The KNT model has been  partly self-criticized~\cite{Krauss:2006eb} because of employing unobservable singlet DM. 
However, recent generalizations~\cite{Ahriche:2014cda, Ahriche:2014oda, Chen:2014ska} of the KNT three-loop topology 
employ non-singlet multiplets which provide charged components as their tracers.
Thereby, while Ref.~\cite{Ahriche:2014cda} still imposes an exact $Z_2$ symmetry for a real fermion triplet DM, Ref.~\cite{Ahriche:2014oda} considers 
a real fermion quintuplet in the context of softly-broken accidental $Z_2$
symmetry and classifies a ``tower" of model possibilities
realized in Refs. \cite{Krauss:2002px,Ahriche:2014cda,Ahriche:2014oda}.

Our model generalizes these studies to different  three-loop topology, proposed  by Aoki, Kanemura and Seto (AKS)~\cite{Aoki:2008av, Aoki:2009vf}.
It involves a second Higgs doublet and provides  another well motivated scenario  for a study of the  2-Higgs Doublet Model (2HDM) recently reviewed in~\cite{Branco:2011iw}.
Our novel three-loop R$\nu$MDM neutrino mass model is presented in Section 2. 
In Section 3 we present some phenomenological signatures of the Majorana quintuplet and the beyond SM (BSM) scalars at the LHC.
We summarize our results in the concluding section.

\section{Three-Loop R$\nu$MDM Model}

 In the original one-loop R$\nu$MDM model~\cite{Cai:2011qr} the fermion quintuplet $\Sigma \sim (5,0)$ has been proposed to provide
its neutral component  $\Sigma^0$ as a DM candidate on account of an accidental $Z_2$ symmetry.
However, a scalar sextuplet field $\Phi$ which closes a neutrino mass diagram in this model also generates a quartic term~\cite{Kumericki:2012bf},
\begin{equation}\label{dim4}
\lambda \Phi^* \Phi^* \Phi H^* + \mathrm{H.c.} \ \ \ , \ \ \ \Phi^* \Phi^* \Phi H^* =
\Phi^{*iabcd} \Phi^{*pqrst} \Phi_{abpqr} H^{*n} \epsilon_{in} \epsilon_{cs} \epsilon_{dt} \ ,
\end{equation}
which breaks the proposed DM-protecting discrete symmetry. 

Similar DM instability controlled by a single parameter has been studied by authors of Ref.~\cite{Ahriche:2014oda}.
In their model with a scalar quintuplet $\Phi \sim (5,-2)$ and a charged  scalar singlet $S^+ \sim (1,2)$  
there is a single $Z_2$-violating term 
\begin{equation}\label{McDon}
\lambda S^+ \Phi^* \Phi \Phi +  \mathrm{H.c.} \ \ \ , \ \ \ \Phi^* \Phi \Phi  =
\Phi^{*abcd} \Phi_{abkl} \Phi_{cdmn}  \epsilon^{km} \epsilon^{ln} \ ,
\end{equation}
leading to an instability of the neutral component of $\Sigma \sim (5,0)$. However, in the limit $\lambda \rightarrow 0 $ 
their model has an accidental  $Z_2$ symmetry: ($\phi, \Sigma$) $\rightarrow$ ($-\phi, -\Sigma$).

In our modified scenario, where  the SM Higgs in Eq.~(\ref{dim4}) or
the charged singlet scalar in Eq.~(\ref{McDon}) are effectively replaced by non-minimal scalar septuplet, the corresponding quartic-interaction term is $Z_2$-even and thus harmless.

\subsection{Structure and Role of 2HD Sector}

\begin{figure}[t]
\begin{center}
\includegraphics[scale=1.0]{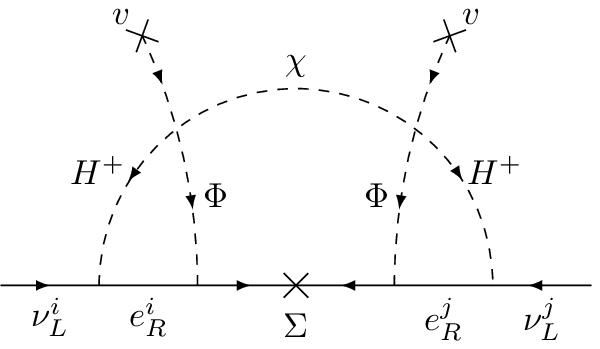} \hspace{0.9cm}
\includegraphics[scale=1.0]{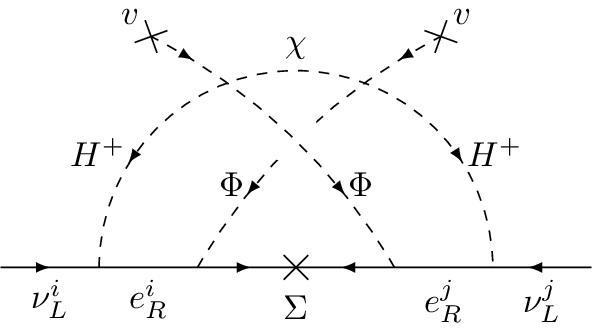}
\caption{\small The diagrams for generating tiny neutrino masses.}
\label{diag-numass}
\end{center}
\end{figure}

The  mass matrix $M^\nu_{ij}$ of active neutrinos is generated by three-loop diagrams in Fig.~\ref{diag-numass} which belong to the AKS topology. 
The outer loops of these diagrams are opened by a pair of charged scalars $H^\pm$ that originates from the two Higgs doublets. 
This  enables the conversion of an active neutrino to the SM lepton singlet $e_R^i$, connecting in the next step the outer neutrino lines 
with the inner one-loop box formed by exotic non-singlet particles.

The two Higgs doublets $H_{\bf{1,2}}\sim (2,1)$ of a generic non-supersymmetric 2HDM,  on which our model is built, can be written as
\begin{equation}
H_{\bf{1}}=\left(\begin{array}{c}
\displaystyle G^+\cos\beta -H^+\sin\beta  \\
\displaystyle \frac{1}{\sqrt{2}}\left(v_1-h\sin\alpha+H\cos\alpha+\mathrm{i}\left( G\cos\beta-A\sin\beta \right)\right)
\end{array}
\right),
\end{equation}
\begin{equation}
H_{\bf{2}}=\left(\begin{array}{c}
\displaystyle G^+\sin\beta +H^+\cos\beta  \\
\displaystyle \frac{1}{\sqrt{2}}\left(v_2+h\cos\alpha+H\sin\alpha+\mathrm{i}\left( G\sin\beta+A\cos\beta \right)\right)
\end{array}
\right),
\end{equation}
where their electroweak vacuum expectation values (VEVs) define $\tan\beta\equiv v_2 /v_1$.
Besides physical charged scalars $H^\pm$, there are three Goldstone bosons ($G, G^\pm$) and three physical neutral scalars:
two CP-even states  $h$ and $H$ with their mixing angle $\alpha$, and a CP-odd neutral scalar $A$.

Let us note that the VEVs $v_1$ and $v_2$ (which are related to the SM VEV $v=$ 246 GeV by $v^2 = v_1^2 + v_2^2$) originate from  $m_{11}^2$ and $m_{22}^2$ terms 
through the minimization conditions of the most general CP-conserving 2HD potential 
\begin{equation}
\begin{split}
V(H_{\bf{1}},H_{\bf{2}})  &= m_{11}^2 H_{\bf{1}}^\dagger H_{\bf{1}}+ m_{22}^2 H_{\bf{2}}^\dagger H_{\bf{2}}
-[m_{12}^2 H_{\bf{1}}^\dagger H_{\bf{2}}+ \, \text{h.c.} ] \\
& +\frac{1}{2}\lambda_1(H_{\bf{1}}^\dagger H_{\bf{1}})^2
+\frac{1}{2}\lambda_2(H_{\bf{2}}^\dagger H_{\bf{2}})^2\\
& +\lambda_3(H_{\bf{1}}^\dagger H_{\bf{1}})(H_{\bf{2}}^\dagger H_{\bf{2}})
+\lambda_4(H_{\bf{1}}^\dagger H_{\bf{2}})(H_{\bf{2}}^\dagger H_{\bf{1}}) \\
& +\left\{\frac{1}{2}\lambda_5(H_{\bf{1}}^\dagger H_{\bf{2}})^2
+\big[\lambda_6(H_{\bf{1}}^\dagger H_{\bf{1}})
+\lambda_7(H_{\bf{2}}^\dagger H_{\bf{2}})\big]
H_{\bf{1}}^\dagger H_{\bf{2}}+\, \text{h.c.}\right\}\,.
\label{2HDpot}
\end{split}
\end{equation}
Here the quartic couplings $\lambda_1$ to $\lambda_5$ can be
traded for the four physical Higgs boson masses as free input parameters and the mixing parameter $\sin(\beta-\alpha)$.

The Yukawa couplings of the fermions are {\it a priori} free parameters, but then they lead to  flavor-changing neutral currents (FCNC) mediated 
by 2HD scalars at the tree level. Usually they are eliminated by imposing a discrete symmetry 
under which ($H_{\bf{1}}, H_{\bf{2}}$) $\rightarrow$ ($H_{\bf{1}}, - H_{\bf{2}}$).
This symmetry  imposed on the potential~(\ref{2HDpot}), henceforth denoted as $\tilde Z_2$, is exact as long as $m_{12}^ 2$,
$\lambda_6$ and $\lambda_7$ vanish.
A recent detailed study  within the 2HD scenario~\cite{Chakrabarty:2014aya} shows that the exact $\tilde Z_2$, in an absence of the soft breaking $m_{12}^ 2$ term, 
does not require intervention of new physics below $\sim$10 TeV scale. Note that at this scale the exotic states of our model already entered into the play.

Out of four different ways the Higgs doublets are conventionally assigned charges under a $\tilde Z_2$ symmetry~\cite{Kanemura:2014bqa}, 
we adopt here the  "lepton-specific" (Type X or Type IV) model 
implemented by AKS~\cite{Aoki:2008av, Aoki:2009vf}, corresponding to Table~\ref{Z-charges}.
In this model $H_{\bf{2}}$ couples to all quarks whereas $H_{\bf{1}}$ couples to all leptons and provides the $H^+ \nu_L e_R$ 
coupling enhanced by $\tan\beta$.

\begin{table}
\begin{center}
  \begin{tabular}{c|ccccc|cc|ccc}
   \hline
   & $Q_i$ & $u_{i R}$ & $d_{i R}$ & $L_{i L}$ & $e_{i R}$ & $H_{\bf{1}}$ & $H_{\bf{2}}$ & $\Phi$ &
    $\chi$ & $\Sigma_{\alpha}$ \\\hline
$Z_2\frac{}{}$                {\rm accidental} & $+$ & $+$ & $+$ & $+$ & $+$ & $+$ & $+$ & $-$ & $-$ & $-$ \\ \hline  
$\tilde{Z}_2\frac{}{}$ {\rm exact,\hspace{1mm}imposed}& $+$ & $-$ & $-$ & $+$ &
                       $+$ & $+$ & $-$ & $+$ & $-$ & $+$ \\\hline
   \end{tabular}
\end{center}
  \caption{Charge assignment under an automatic $Z_2$ and an imposed $\tilde Z_2$ symmetry in the model.}
  \label{Z-charges}
 \end{table}

\subsection{Exotic BSM Multiplets}

The model that we propose is  based on the symmetry of the SM gauge group $SU(3)_C \times SU(2)_L \times U(1)_Y$, where in the following only the relevant 
electroweak part is explicated.  In addition to usual SM fermions, we introduce three generations of exotic real fermions  
transforming as $\Sigma_{\alpha} \sim (5,0)$, where $\alpha =1,2,3$ labels generations. 
Also, in addition to the described scalar doublets there are two exotic scalars $\Phi \sim (5,-2)$ and $\chi \sim (7,0)$.
In Table~\ref{Z-charges} we summarize the assignment of the charges under an imposed $\tilde Z_2$ symmetry and by it guaranteed accidental exact $Z_2$ symmetry in our model.
All exotic additional fields are totally symmetric tensors $\Sigma_{abcd}$, $\Phi_{abcd}$  and $\chi_{abcdef}$, and 
the components of $\Sigma_{abcd}$ read
\begin{eqnarray}
    \begin{matrix} \Sigma_{1111} = \Sigma_R^{++} \\ \Sigma_{1112} = \frac{1}{\sqrt{4}}\Sigma_R^{+} \\ \Sigma_{1122} =  \frac{1}{\sqrt{6}}\Sigma_R^{0} \\ 
    \Sigma_{1222} = \frac{1}{\sqrt{4}}(\Sigma_L^{+})^c  \\ \Sigma_{2222} =  (\Sigma_L^{++})^c   \end{matrix}  \ \ ,
\end{eqnarray}
and form two charged Dirac fermions and one neutral Majorana fermion
\begin{equation}\label{fermions}
   \Sigma^{++} = \Sigma^{++}_R + \Sigma^{--C}_R\ ,\ \Sigma^+ = \Sigma^+_R - \Sigma^{-C}_R\ ,\ \Sigma^0
= \Sigma^0_R + \Sigma^{0C}_R\ .
\end{equation}
The components of two exotic scalar fields read 
\begin{eqnarray}
    \begin{matrix} \Phi_{1111} = \phi^+ \\ \Phi_{1112} = \frac{1}{\sqrt{4}}\phi^{0} \\ \Phi_{1122} =  \frac{1}{\sqrt{6}}\phi^{-} \\ 
    \Phi_{1222} = \frac{1}{\sqrt{4}}\phi^{--}  \\ \Phi_{2222} =  \phi^{---}   \end{matrix}   \ \ ,
&& \begin{matrix} \chi_{111111} = \chi^{+++} \\ \chi_{211111} = 
    \frac{1}{\sqrt{6}}\chi^{++} \\ \chi_{221111} = \frac{1}{\sqrt{15}}\chi^{+}\\ 
    \chi_{222111} = \frac{1}{2\sqrt{5}}\chi^{0}  \\ \chi_{222211} = 
    \frac{1}{\sqrt{15}}\chi^- \\ \chi_{222221} = 
\frac{1}{\sqrt{6}}\chi^{--} \\ \chi_{222222} = \chi^{---} \end{matrix} \ \ ,
\end{eqnarray}
where we distinguish $\phi^{-}$ and $(\phi^{+})^*$.
The Yukawa interaction is given by 
\begin{eqnarray}\label{yukawa}
 {\cal L}_Y
   = - y_{e_i}  \overline{L}_{i L} H_{\bf{1}} e_{i R}
     - Y_{i \alpha}  \overline{(e_{i R})^c} {\Phi}^* \Sigma_{\alpha R} + \mathrm{h.c.} \ ,
\end{eqnarray}
where only the Higgs doublet $H_{\bf{1}}$ in the lepton-specific 2HD model couples to SM leptons. 
Therefore the SM lepton mass $m_e$, which in ``minimal" SM corresponds to the Yukawa strength $y_{e_i}^{SM}
=\sqrt{2}m_{e_i}/v$, in the 2HD context reads $\sqrt{2}m_{e_i} \mathrm{tan}\beta/v$. 
The Yukawa terms involving new fields read in components as
\begin{eqnarray}\label{lagrangian}
\nonumber  \big(\overline{\Sigma_R}\big)^{klmn}  \Phi_{klmn}  (e_{R})^c &=& 
    \phi^{---} \overline{\Sigma_R^{--}}  (e_{R})^c  + \phi^{--} \overline{\Sigma_R^{-}}  (e_{R})^c +  \phi^{-} \overline{\Sigma_R^{0}}  (e_{R})^c   \\
    &+&  \phi^{0} \overline{\Sigma_R^{+}}  (e_{R})^c  +  \phi^{+} \overline{\Sigma_R^{++}}  (e_{R})^c      \ .
\end{eqnarray}
The scalar potential contains the gauge invariant pieces 
\begin{eqnarray}\label{scalarpot}
\nonumber  V(H_{\bf{1}},H_{\bf{2}},\Phi, \chi) &=& V(H_{\bf{1}},H_{\bf{2}}) + V(\Phi)  + V(\chi)  \\ 
\nonumber   &+&  V_m(H_{\bf{1}},H_{\bf{2}},\Phi) + V_m(H_{\bf{1}},H_{\bf{2}},\chi) + V_m(\Phi,\chi)   \\
            &+&  V_m(H_{\bf{1}},H_{\bf{2}},\Phi, \chi)  \ ,
\end{eqnarray}
where $V(H_{\bf{1}},H_{\bf{2}})$ is given in Eq.\,(\ref{2HDpot}). The $\tilde Z_2$-symmetric mixing potential $V_m(H_{\bf{1}},H_{\bf{2}},\Phi, \chi)$ 
provides the quartic term
\begin{eqnarray}\label{NEWdim4}
\nonumber   V_m(H_{\bf{1}},H_{\bf{2}},\Phi, \chi) &=&  \kappa H_{\bf{1}} H_{\bf{2}} \Phi \chi + \mathrm{h.c.} \ ,\\ 
    H_{\bf{1}} H_{\bf{2}} \Phi \chi &=&  H_{\bf{1}\it{i}} H_{\bf{2}\it{j}} \Phi_{klmn}  \chi_{abcdfg} \epsilon^{ng} \epsilon^{ia} \epsilon^{jb} \epsilon^{kc} \epsilon^{ld} \epsilon^{mf}\ ,
\end{eqnarray}
which includes the couplings needed to close our three-loop mass diagrams: 
\begin{eqnarray}\label{quarticHHPhichi}
    \nonumber H_{\bf{1}} H_{\bf{2}} \Phi \chi  \supset - \frac{1}{\sqrt{6}}  \chi^{--} H^{+}_{\bf{1}} H^{0}_{\bf{2}} \phi^{+} + \frac{2}{\sqrt{15}} \chi^{-} H^{+}_{\bf{1}} H^{0}_{\bf{2}} \phi^{0} 
    - \frac{\sqrt{3}}{\sqrt{10}} \chi^{0}  H^{+}_{\bf{1}} H^{0}_{\bf{2}} \phi^{-} \\
    + \frac{2}{\sqrt{15}} \chi^{+} H^{+}_{\bf{1}} H^{0}_{\bf{2}} \phi^{--}  - \frac{1}{\sqrt{6}} \chi^{++}  H^{+}_{\bf{1}} H^{0}_{\bf{2}} \phi^{---} + (H_{\bf{1}} \leftrightarrow  H_{\bf{2}}) \ .
 \end{eqnarray}
These couplings enable the decays of the exotic scalars, making the scalar septuplet unstable.

Let us stress that without an important interplay of the 2HD and exotic sectors in our model
there would be an additional dimension-three $Z_2$-noninvariant operator
\begin{eqnarray}\label{NEWdim3}
  \mu \Phi\Phi^* \chi &=& \mu \Phi_{ijkl} \Phi^{*iabc} \chi_{abcpqr} \epsilon^{jp} \epsilon^{kq} \epsilon^{lr} \ .
\end{eqnarray}
This coupling, which could make the scalar septuplet DM candidate $\chi^0$ unstable already at the tree level and which also threatens a stability 
of the fermion quintuplet, would be a single $Z_2$-noninvariant term\footnote{Another potentially dangerous trilinear term  $\chi^3$ 
is Bose-forbidden for a septuplet field $\chi$.} in our model. However, the operator in~(\ref{NEWdim3}) is forbidden 
by the $\tilde Z_2$ symmetry enforced on the 2HD sector and mandatory for a whole model, with parities explicated in Table~\ref{Z-charges}.

\subsection{Three-Loop-Induced Neutrino Mass}

The neutrino mass term 
\begin{eqnarray}
  {\cal L}^{\rm eff} = \overline{\nu_L^c}_i M_{ij} {\nu_L}_j
 \end{eqnarray}
is generated by the three-loop diagrams in Fig.~\ref{diag-numass}.
They include five diagrams corresponding to five pairs of ($\chi$, $\Sigma$) fields 
propagating in the inner loop in Fig.~\ref{diag-numass},
\begin{eqnarray} 
(\chi^{--},\Sigma^{++}),\; (\chi^{-},\Sigma^{+}),\;(\chi^{0},\Sigma^{0}),\;
(\chi^{+},(\Sigma^{+})^c),\;(\chi^{++},(\Sigma^{++})^c)   \;  .
\end{eqnarray}
They correspond to five components of the $\Phi$ field in the loop ($\phi^+$ to $\phi^{---}$) building the quartic vertices which result from 
the following substitution in five terms listed in Eq.~(\ref{quarticHHPhichi}):
\begin{eqnarray}\label{HHPhichi-vertex}
H^{+}_{\bf{1}} H^{0}_{\bf{2}} +  H^{+}_{\bf{2}} H^{0}_{\bf{1}} \,  \rightarrow  \, v\,  \mathrm{cos}2\beta\,   H^{+}\ .
\end{eqnarray}
In conjunction with appropriate Yukawa couplings, these quartic couplings lead to radiatively generated
lepton number breaking Majorana neutrino masses.\\
When we neglect the mass differences within $\Phi, \chi$ and $\Sigma_{\alpha}$  multiplets, the mass matrix $M^\nu_{ij}$ for active neutrinos keeps the AKS form \cite{Aoki:2009vf}
\begin{eqnarray}
M_{ij} &=& \sum_{\alpha=1}^3 
   C_{ij}^\alpha \, F(m_{H^\pm}^{},m_\Phi^{}, m_\chi, m_{\Sigma_{\alpha}}), \label{eq:mij}
\end{eqnarray}
where the coefficient $C_{ij}^\alpha$ comprises the vertex coupling strengths
\begin{eqnarray}
C_{ij}^\alpha &=&
    \frac{7}{3} \kappa^2  \tan^2\beta \cos^2 2\beta \,
  y_{e_i}^{\rm SM} Y_i^\alpha y_{e_j}^{\rm SM} Y_j^\alpha, 
\end{eqnarray}
and the loop integral is represented by function $F$,
\begin{multline}
\!\!\!\!\! F(m_{H^\pm}^{},m_\Phi^{}, m_\chi, m_{\Sigma_{\alpha}}) =
 \left(\frac{1}{16\pi^2}\right)^3 \frac{(-m_{\Sigma_{\alpha}}^{})}{m_{\Sigma_{\alpha}}^2-m_\chi^2} \,
 \frac{v^2}{m_{H^\pm}^4}  \\
  \times \int_0^{\infty} \!\! x dx
  \left\{B_1(-x,m_{H^\pm}^{},m_\Phi^{})-B_1(-x,0,m_\Phi^{})\right\}^2
  \left(\frac{m_{\Sigma_{\alpha}}^2}{x+m_{\Sigma_{\alpha}}^2}-\frac{m_\chi^2}{x+m_\chi^2}\right)\nonumber \, ,
\end{multline}
where $B_1$ denotes the Passarino-Veltman function
for one-loop integrals~\cite{passarino-veltman}.

\begin{figure}[t]
\centerline{\includegraphics[scale=0.6]{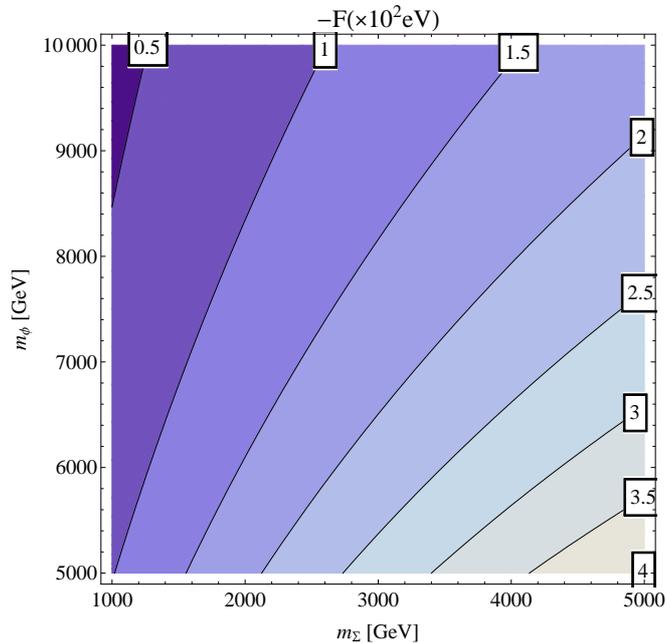}}
\caption{\small The contour plot of the loop-integral function 
    $-F (\times 10^{2}\,{\rm eV})$ 
    that practically does not change for $M_{H^\pm}$ between 100 GeV to 1 TeV
and $m_\chi \geq m_\Phi^{}$ in 5 to 20 TeV range.}
\label{nu_valueF}
\end{figure}

The magnitude of the integral function $F$ in Fig.~\ref{nu_valueF} is practically insensitive 
on the mass of the charged Higgs boson $H^\pm$ in the 100 GeV to 1 TeV range discussed in Sect. 3.3.2. Its magnitude is plotted 
as a function of $m_\Sigma $ for the values of $m_\Sigma \leq m_\Phi^{} $. It is also rather insensitive on the value of $m_\chi$, 
which we take $m_\chi \geq m_\Phi^{}$.
The magnitude of $F$ which is of order $10^{2}$ eV
in the wide range of the parameter space reproduces the neutrino masses  with the coefficient  $C_{ij}^\alpha \leq 10^{-4}$ that is easily 
achieved without fine tuning.

\section{Phenomenology of BSM States}

There are two major phenomenological issues generally related to the models for radiative neutrino masses:
(i) an enlarged Higgs sector with charged scalar bosons;
(ii) the TeV-scale right-handed neutrinos with Majorana masses, as possible DM candidates.
In order to account for all of the observed DM abundance, the mass $m_\Sigma$ of MDM  Majorana quintuplet~\cite{Cirelli:2005uq} approaches 10 TeV range. Accordingly, 
the Majorana quintuplet may be studied at the LHC  only by relaxing this condition like in presence of non-thermal production mechanisms, in non-standard cosmological
scenarios or if it accounts only for a fraction of the DM abundance.

\subsection{DM candidate at the LHC}

The present study of a Majorana quintuplet on colliders can be  compared with a previous one~\cite{Franceschini:2008pz}  of the  Majorana triplet  employed 
in Type-III seesaw. Let us note that the quintuplet in presence of a half integer scalar quartet~\cite{Kumericki:2012bh} $\Phi \sim (4,1)$ leads to neutrino masses both 
at tree level\footnote{Therefore the model of Ref.~\cite{Kumericki:2012bh} has been dubbed~\cite{Law:2013gma} the Type-V seesaw.}
and one-loop level. It has been shown in~\cite{Kumericki:2012bh} that it is possible to falsify this tree-level option at the LHC by mere non-observation of related
light quintuplet states.

Here we are reassessing the LHC production of a quintuplet fermions studied
previously~\cite{Kumericki:2012bh} for the 2011/2012 run of the LHC.  Now we
consider it at a design 14 TeV energy both for the design luminosity of 300 fb$^{-1}$
and for high luminosity (HL-LHC) of 3 ab$^{-1}$, as well as for a futuristic 100 TeV
$\rm pp$ collider with 3 ab$^{-1}$.

\begin{figure}[t]
\begin{center}
\includegraphics[scale=0.9]{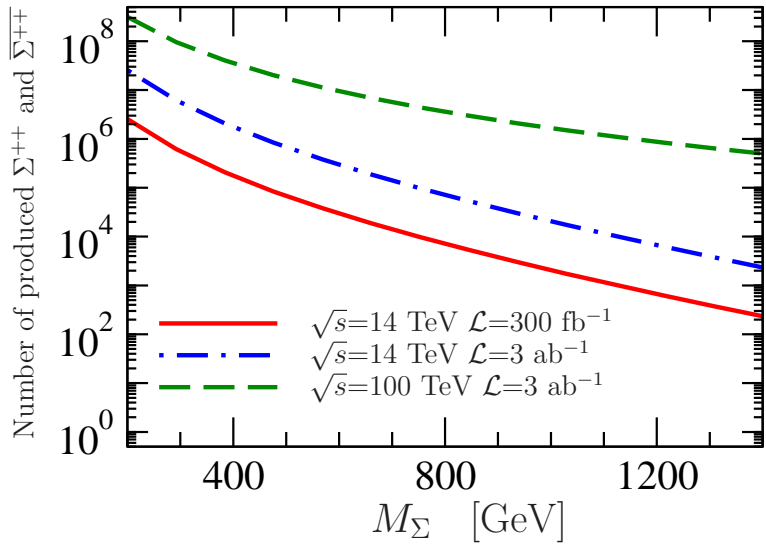}
\caption{Number of $\Sigma^{++}$ and
$\overline{\Sigma^{++}}$ particles produced for three
characteristic collider setups, in dependence on
the heavy lepton mass $M_{\Sigma}$.}
\label{fig:LHCproduction}
\end{center}
\end{figure}

Quintuplet lepton pairs in proton-proton collisions are mostly produced via
Drell-Yan process, mediated by neutral and charged gauge bosons. Detailed
expressions for cross-sections are given in~\cite{Kumericki:2012bh}. 
In particular, for $M_\Sigma = 400 \, \rm{GeV}$ and $300\,{\rm fb}^{-1}$ of
integrated luminosity at $\sqrt{s}=14$ TeV, the LHC should produce about $1.8\cdot
10^5$ doubly-charged $\Sigma^{++}$ or $\overline{\Sigma^{++}}$ fermions,
and $2.2\cdot 10^5$ $\Sigma - \overline{\Sigma}$ pairs in total.
In Fig.~\ref{fig:LHCproduction} we plot the expected number of
produced $\Sigma^{++}$ and $\overline{\Sigma^{++}}$ particles for three
characteristic collider setups in dependence on the heavy lepton mass $M_\Sigma$.
There would be roughly the same number of singly-charged $\Sigma$ fermions produced, 
and half as many neutral ones.

The mass differences within the multiplet are of the order of few hundred MeV, so
charged $\Sigma$ leptons decay to neutral DM candidate by cascade radiation of soft
undetected pions, e.g. $\Sigma^{++}\to \Sigma^{+}\pi^+ \to \Sigma^0 \pi^+ \pi^+$.

Detailed study of collider signature of wino-like DM triplet in 
\cite{Cirelli:2014dsa} identified disappearing tracks and monojets as
most promising search channels. It turns out that regarding monojets
our model is similar enough that we can, within some reasonable
assumptions, directly use properly transformed results of \cite{Cirelli:2014dsa}.
In particular, we have monojets generated by same diagrams as in the
triplet model, and, additionally, diagrams with doubly-charged leptons.
Taking into account these additional contributions and different
electroweak charges, in our model five times as many DM particles are produced
as in the triplet model; either directly or via cascade decays mentioned above.

Since this enhancement factor turns out to be the same for various production
mechanisms  any kinematic cuts 
will cut away the same fraction of signal in both models\footnote{Neutral current 
partonic production processes are enhanced
by electric charge factors (thanks to the hypercharge being zero) like
$Q(\Sigma^{++})^2 + Q(\Sigma^{+})^2 = 2^2 + 1^2 = 5$,
while the  corresponding charged current partonic production
processes are enhanced by electroweak SU(2) ladder operators
matrix elements $\sqrt{(T \mp T_3)(T \pm T_3 + 1)/2}$, giving
$\sqrt{2}^2 + \sqrt{3}^2 = 5$ again.}, and SM background
should also be about the same. Thus we can
estimate the LHC monojet search reach for our model by taking
significance $Z_{\rm triplet}$ as defined in Eq. (3.1) of \cite{Cirelli:2014dsa}
and plotted in their Fig. 2, and scaling it using formula
\begin{equation}
    Z_{\rm quintuplet} = \frac{1}{
        \sqrt{ \left(\frac{\sigma_{\rm triplet}}{\sigma_{\rm quintuplet}}\right)^2  
        \left(\frac{1}{Z_{\rm triplet}^2} - \beta^2\right) + \beta^2}} \;,
\label{eq:Z} 
\end{equation}
where $\sigma$ is the total cross section for Drell-Yan production of $\Sigma$
lepton pair with any charges, and $\beta$ is the signal systematic uncertainty, which we
should take to be 10\%, same as in \cite{Cirelli:2014dsa}.

\begin{figure}[t]
\centerline{\includegraphics[scale=0.6]{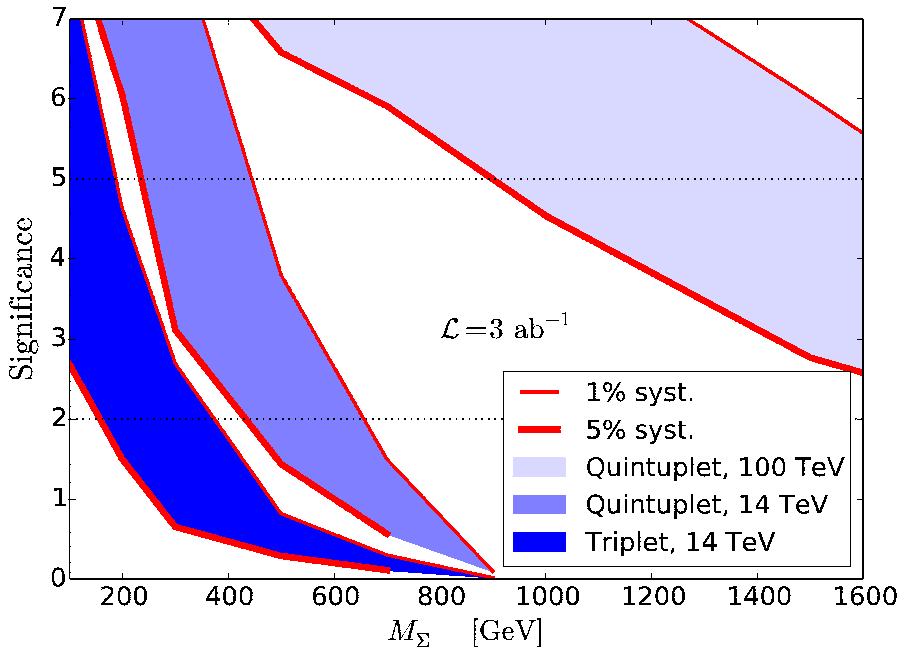}}
\caption{Reach of monojet searches for quintuplet DM at the high-luminosity LHC and at
a futuristic 100 TeV proton-proton collider. 
Bands correspond to variation of background systematic uncertainty from 5\% to 1\%.
For comparison, corresponding
reach of monojet searches for triplet model from \protect\cite{Cirelli:2014dsa} is
also shown.}
\label{fig:monojet}
\end{figure}

Resulting significance is plotted in Fig.~\ref{fig:monojet}. One notices that whereas
the triplet model can only be \emph{excluded} with 95\% C.L. for $M < 350\, (150)$ GeV, 
the high-luminosity LHC can make a 5$\sigma$ \emph{discovery} of quintuplet model DM
for $M < 450\: (250)$ GeV, with systematic uncertainty of the background
of 1\% (5\%),  respectively. Futuristic 100 TeV collider can easily extend the discovery reach 
beyond 1 TeV.

\subsection{Majorana Quintuplet in the Broken \boldmath$Z_2$ Scenario}

In the proposed model $\tilde{Z}_{2}$ symmetry, introduced in 2HDM for
flavour physics reasons, implies also accidental $Z_2$ symmetry, thus
protecting the stability of $\Sigma^{0}$ as a DM candidate.
However, it is known that soft breaking of $\tilde Z_2$ in 2HDM may have some interesting
features, like introducing a new source of CP violation and preserving stability and
unitarity of the scalar potential.  In the 2HD sector there is a potential soft
$m_{12}$ term~(\ref{2HDpot}), whereas the $Z_2$-breaking soft
term~(\ref{NEWdim3}) in the sector of exotic scalar fields would destabilize
the  Majorana quintuplet $\Sigma^0$ via loop-induced effects.

Let us for completeness estimate the value of the small, 'tHooft-natural
coefficient carried by  the term~(\ref{NEWdim3}),  so that the loop-induced
decays of $\Sigma^0$ state may be long enough in comparison to the lifetime
of the Universe.

\begin{figure}[t]
\centerline{\includegraphics[scale=0.5]{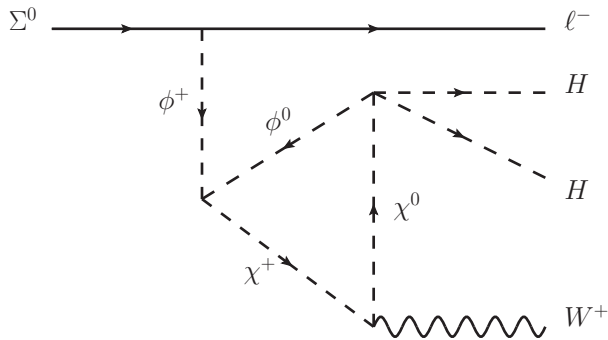}}
\caption{\small An example of decay of the heavy lepton $\Sigma$ at the loop level
in the case of broken $Z_2$ symmetry.}
\label{S0decay}
\end{figure}

For example, the loop diagram in Fig.~\ref{S0decay} leads to the decay amplitude 
for  $\Sigma^0 \rightarrow l_R^- W^+ H^{0}_{\bf{1}} H^{0}_{\bf{2}}$. It can be compared to the decay amplitude for  $\Sigma^0 \rightarrow \nu W^+ W^- H$
in Fig. 1 in~\cite{Kumericki:2012bf}, induced by dim-4 operator which destabilized DM candidate in one-loop R$\nu$MDM model~\cite{Cai:2011qr}. In the present case the width for decay
into the four particle final state $l_R^- W^+ H^{0}_{\bf{1}} H^{0}_{\bf{2}}$ is given by
\begin{equation}
    \Gamma \sim \frac{\kappa^2 g^2}{192 \pi} \Big(\frac{1}{16\pi^2}\Big)^4 Y^2 \mu^2 \frac{m_\Sigma^9}{m_\Phi^8 m^2_W } \, .
\end{equation}
By assuming $\kappa=g=0.65$ and by adopting the values $m_\Sigma=m_\Phi=10 \ \rm{TeV}$ and $Y=10^{-1}$, as used in \cite{Cai:2011qr}, the  lifetime of our DM candidate exceeds 
the age of the Universe $\approx 10^{17} {\rm s}$ if the soft term coupling $\mu$ does not exceed the value of neutrino masses, $\mu < 0.1$ eV.
A stronger bound $\mu < 10^{-9}$ eV can be obtained
in the context of the decaying DM \cite{Boucenna:2012rc,Boucenna:2014zba,Audren:2014bca}.
Notably, switching off the soft $\mu$ term does not affect the neutrino mass
diagram but makes the $\Sigma^0$ state a viable DM candidate.

So, in order to confirm the relation of the quintuplet particles to neutrinos
one has to study their decays.  Let us stress that there are ample decays to
purely SM final state particles in case that the quintuplets are not
constrained by $Z_2$ symmetry, like in scenarios where they are accompanied by
even-plet scalars~\cite{Kumericki:2012bh,Cai:2011qr}.
 
Therefore a recent scrutiny~\cite{Yu:2015pwa} is welcome as a way to possibly
exclude such modes and scenarios, what would be in favor of the present
three-loop model in which fermion quintuplets are accompanied by odd-plet
scalars. Then only less visible cascade decays of charged $\Sigma$ states
remain, and the search is focused on monojets or disappearing tracks like in 
case of wino-like particles~\cite{Cirelli:2014dsa}.  In such scenario the
$\Sigma^0$ is the DM candidate.

\subsection{BSM Scalars at Colliders}

The 2HD scalar sector has been studied in detail (for a review see~\cite{Branco:2011iw} and~\cite{Kanemura:2014bqa}), 
independently of possible further extensions of the scalar sector. A study in the context of additional singlet scalars of AKS model 
faces challenging separation of signals addressed in~\cite{Aoki:2010tf}.
The case of the larger multiplets of the present account is characterized by the fact that 
the 2HD and the exotic scalar sectors are decoupled due to the SM gauge symmetry, and thus can be treated separately.

\subsubsection{Exotic Quintuplet and Septet Scalars}

Since in this work the scalar quintuplet and septuplet states are assumed to be heavier than Majorana quintuplet accounting for all DM abundance, 
we do not expect that they are accessible at the LHC. In particular, the collider signatures of these exotic scalars are out of scope of the present paper.
Still, their multiple-charged components are studied in number of papers for different reason.

As already pointed out, the sectors of small and large scalar multiplets are to large extent decoupled and can be treated separately.
Here we are explicating the $Z_2$ symmetry breaking terms connecting small and large scalar multiplets, which are allowed by  the exact $\tilde Z_2$ symmetry.
A doublet-quintuplet mixing is represented by a dim 5 operator, 
\begin{equation}
\frac{1}{\Lambda} \Phi \Phi^* \Phi H_{\bf{1}} H_{\bf{1}}  \;, \quad \frac{1}{\Lambda} \Phi \Phi^* \Phi H_{\bf{2}} H_{\bf{2}}  \;. 
\end{equation}
However, in order to enable similar doublet-septuplet mixing,  we must go to a suppressed dim 7 operator
\begin{equation}
\frac{1}{\Lambda^3} \chi  ( H_{\bf{1}} H_{\bf{2}}^* )^3 \;.
\end{equation}
Since the natural mass scale of these fields is in a multi-TeV range, a study of 2HD states may be more promising.

\subsubsection{Charged Higgs in Lepton-specific 2HDM}

The charged Higgs boson phenomenology for lepton-specific 2HDM at LEP was presented in~\cite{Logan:2009uf}, and for direct searches at LHC and ILC in~\cite{Kanemura:2014dea}.
The standard procedure~\cite{Kajiyama:2013sza,Kanemura:2014hja,Biswas:2014uba} is to adjust the scenario for the neutral $h$ state to be the SM-like Higgs boson with mass 125 GeV.
In the first case corresponding to the {\it decoupling regime} where $\sin(\beta-\alpha) \simeq 1$, the $H^\pm$ mass $M\gg v$ may be related to large soft breaking scale of the discrete symmetry $\tilde Z_2$.
In another case where the $H^\pm$ mass is at the scale $M\sim v$, 
these charged states may be probed through the Higgs diphoton decay.

Let us note that there is another three-loop model~\cite{Kajiyama:2013lja} which has a lepton-specific 2HDM as a part of the scalar sector extension. 
Therefore the Higgs phenomenology presented there applies to the charged states $H^\pm$. 
On the other hand for a charged Higgs heavier than top, the BaBar data based on $b \rightarrow s \gamma$ exclude a charged Higgs
lighter than 380 GeV, independently of $\tan\beta$ (see Ref.~\cite{Hashemi:2014oka} and references therein).
In addition, in Ref.~\cite{Enberg:2014pua} the charged Higgs decay channels have been identified, which are promising for non-supersymmetric 2HD models.

\section{Conclusions}

We propose a model addressing in a common framework the open questions of the neutrino masses and the dark matter of the Universe.
We follow  the idea of the R$\nu$MDM  proposed at one-loop level in~\cite{Cai:2011qr} and look at alternative particle content that may actually realize it at three-loop level.
We adopt a three-loop topology proposed by AKS~\cite{Aoki:2008av, Aoki:2009vf} in a way that their imposed  $Z_2$ symmetry 
appears accidentally due to higher representations added to the SM particle content. 
Thereby we keep a $Z_2$-even second Higgs doublet field, while the $Z_2$-odd weak singlet fields $S$, $\eta$ and $N_R$ of AKS 
are replaced by our higher multiplets $\Phi$, $\chi$ and $\Sigma$, respectively.

We can compare the present study to the recent~\cite{Aoki:2011zg, Dev:2012sg, Queiroz:2014pra} and in particular to~\cite{Ahriche:2014cda, Ahriche:2014oda} which are based on 
modifications of KNT topology~\cite{Krauss:2002px} and include also higher multiplets.
A common feature of the models in~\cite{Ahriche:2014cda, Ahriche:2014oda} are super-renormalizable terms violating a
desired $Z_2$ symmetry.  In our case such
unique term~(\ref{NEWdim3}) is forbidden by exact $\tilde Z_2$ symmetry  originating in the 2HD sector in our setup.
The $\tilde Z_2$ charges have to be attributed also to exotic sector states as displayed in Table~\ref{Z-charges},
in order to enable the neutrino mass loop diagram.
The essential couplings in Eq.~(\ref{quarticHHPhichi}) make a neutral component of the scalar septuplet unstable so that 
the Majorana quintuplet $\Sigma \sim (5,0)$ remains a single DM candidate. 

There are cosmological scenarios in which this DM candidate can be within the
reach of the LHC, where we take its mass as a free parameter.
The studies of testability of the exotic quintuplet states at hand that are testable at the LHC can be compared
to the LHC phenomenology of wino-like fermion triplet~\cite{Cirelli:2014dsa}
belonging to recent minimalistic variant of MDM model.
Their phenomenology reduces to a restricted subset of decays which are allowed by a $Z_2$ symmetry. Mere observation of a decays into purely SM final state particles 
would falsify the  scenario with exact $Z_2$ symmetry. 
In the absence of such decays the proposed set of exotic BSM particles keeps its appealing features  of contributing to the DM and generating three-loop suppressed neutrino masses.
Using monojet searches, our DM candidate with mass $M < 450\,{\rm GeV}$
is discoverable at the high-luminosity LHC.

\subsubsection*{Acknowledgment}
We thank Branimir~Radov\v{c}i\'c for useful discussions at the early stages of
this work. We also thank the authors of ~\cite{Cirelli:2014dsa} for sharing numbers
corresponding to the their monojet analysis plotted in our
Fig.~\ref{fig:monojet}.
This work has been supported in part by the Croatian Science Foundation under
the project number 8799.

\end{document}